\newcommand{\etatil}{\tilde{\eta}}
\newcommand{\rtsix}{{1\over\sqrt6}}
\newcommand{\third}{{\ts{1\over 3}}}
\newcommand{\sixth}{{\ts{1\over 6}}}
\newcommand{\ninth}{{\ts {1\over 9}}}
\newcommand{\thaa}{\theta}
\newcommand{\thaap}{{\theta^\prime}}
\newcommand{\phii}{\varphi}
\newcommand{\phiip}{{\varphi^\prime}}
\newcommand{\lpr}{l^\prime}
\newcommand{\mpr}{m^\prime}
\newcommand{\rpr}{r^\prime}
\newcommand{\ie}{{\it i.e.}}
\newcommand{\eg}{{\it e.g.}}

\newcommand{\be}{\begin{equation}}
\newcommand{\ee}{\end{equation}}
\newcommand{\bea}{\begin{eqnarray}}
\newcommand{\eea}{\end{eqnarray}}
\newcommand{\ba}{\begin{array}}
\newcommand{\ea}{\end{array}}
\newcommand{\al}{\alpha}
\newcommand{\ga}{\gamma}
\newcommand{\de}{\delta}

\newcommand{\om}{\omega}
\newcommand{\ep}{\epsilon}
\newcommand{\tha}{\theta}

\newcommand{\la}{\lambda}
\newcommand{\La}{\Lambda}
\newcommand{\Om}{\Omega}

\newcommand{\ka}{\kappa}
\newcommand{\del}{\mbox{\raisebox{2pt}{$\bigtriangledown$}}}
\newcommand{\tr}{\mathrm{tr}}

\newcommand{\R}{\mathrm{I}\kern -2.5pt \mathrm{R}}
\newcommand{\Z}{\mathsf{Z}\kern -5pt \mathsf{Z}}
\newcommand{\C}{\mathsf{I}\kern -5pt \mathrm{C}}

\newcommand{\ua}{\underline{a}}
\newcommand{\ub}{\underline{b}}

\newcommand{\D}{{\rm d}}
\newcommand{\pa}{\partial}
\newcommand{\rar}{\rightarrow}
\newcommand{\non}{\nonumber}
\newcommand{\we}{\wedge}
\newcommand{\cN}{\mathcal{N}}
\newcommand{\cR}{\mathcal{R}}
\newcommand{\cO}{\mathcal{O}}

\newcommand{\half}{\mbox{$\frac{1}{2}$}}

\newcommand{\ts}{\textstyle}

\newcommand{\1}{1\kern -3pt \mathrm{l}}

\newcommand{\SU}{\mathrm{SU}}
\newcommand{\SO}{\mathrm{SO}}
\newcommand{\Sp}{\mathrm{Sp}}

\newcommand{\U}{\mathrm{U}}

\newcommand{\e}{{\rm e}}

\newcommand{\quarter}{\mbox{$\frac{1}{4}$}}
\newcommand{\UR}{\U(1)_{\cR}}
\newcommand{\Ur}{\U(1)_{R}}

\documentclass[12pt]{article}

\newdimen\tableauside\tableauside=1.0ex
\newdimen\tableaurule\tableaurule=0.4pt
\newdimen\tableaustep
\def\phantomhrule#1{\hbox{\vbox to0pt{\hrule height\tableaurule
width#1\vss}}}
\def\phantomvrule#1{\vbox{\hbox to0pt{\vrule width\tableaurule
height#1\hss}}}
\def\sqr{\vbox{%
  \phantomhrule\tableaustep

\hbox{\phantomvrule\tableaustep\kern\tableaustep\phantomvrule\tableaustep}%
  \hbox{\vbox{\phantomhrule\tableauside}\kern-\tableaurule}}}
\def\squares#1{\hbox{\count0=#1\noindent\loop\sqr
  \advance\count0 by-1 \ifnum\count0>0\repeat}}
\def\tableau#1{\vcenter{\offinterlineskip
  \tableaustep=\tableauside\advance\tableaustep by-\tableaurule
  \kern\normallineskip\hbox
    {\kern\normallineskip\vbox
      {\gettableau#1 0 }%
     \kern\normallineskip\kern\tableaurule}%
  \kern\normallineskip\kern\tableaurule}}
\def\gettableau#1 {\ifnum#1=0\let\next=\null\else
  \squares{#1}\let\next=\gettableau\fi\next}

\newcommand{\Yfund}{\tableau{1}}
\newcommand{\Ysymm}{\tableau{2}}
\newcommand{\Yasymm}{\tableau{1 1}}

\begin{document}

\begin{flushright}
BRX-TH-492 \\
HUTP-01/A028 \\
BOW-PH-122  \\
{\tt hep-th/0106020}\\
\end{flushright}
\vspace{1mm}
\begin{center}
{\bf\Large\sf
${1}/{N}$ corrections to anomalies and the {\large AdS/CFT} correspondence \\
for orientifolded {\large $\cN=2$} orbifold and {\large $\cN=1$}
conifold models}
\end{center}
\vskip 5mm
\begin{center}

Stephen G. Naculich\footnote{
Research supported in part by the NSF under grant no.~PHY94-07194 \\
\phantom{aaa}  through the ITP Scholars Program.}$^{,a}$,
Howard J. Schnitzer
\footnote{Research supported in part by the DOE under grant
DE--FG02--92ER40706.}
${}^{\!\!\!,\!\!\!}$
\footnote{Permanent address.}$^{,b,c}$,
and Niclas Wyllard\footnote{
Research supported by the DOE under grant DE--FG02--92ER40706.\\
{\tt \phantom{aaa} naculich@bowdoin.edu;
schnitzer,wyllard@brandeis.edu}\\}$^{,b}$
\end{center}
\vspace{.15in}

\begin{center}
$^{a}${\em Department of Physics\\
Bowdoin College, Brunswick, ME 04011}

\vspace{.2in}

$^{b,3}${\em Martin Fisher School of Physics\\
Brandeis University, Waltham, MA 02454}

\vspace{.2in}

$^{c}${\em Lyman Laboratory of Physics \\
Harvard University, Cambridge, MA 02138}
\end{center}
\vspace{.3in}

\vskip 5mm

\begin{abstract}
We show that for a large class of
$d=4$ $\cN=2$ conformal field theories
the $1/N$ correction to the chiral anomaly of the $\U(1)$ R-current
can be shown to arise from the D7-branes present in the dual
orientifolded orbifold string theories,
generalizing a result in the literature for the simplest case.
We also study the $\U(1)_R$ anomaly for
 $d=4$ $\cN=1$ conformal field theories
that arise from orientifolds of the conifold.
We find agreement between the field- and string-theoretic
calculations, confirming a prediction of the AdS/CFT correspondence
at order $1/N$ for string theories on $\mathit{AdS}_5{\times}\, T^{11}/\Z_2$.
\end{abstract}

\newpage

\setcounter{equation}{0}
\section{Introduction}
There exists considerable evidence supporting the AdS/CFT
correspondence \cite{Maldacena:1998}
in the limit of large $N$ \cite{Aharony:1999b},
but the $1/N$ corrections to the
leading large-$N$ results are still rather poorly understood.
Corrections of order $1/N^2$ are difficult to study, as they involve
string loop corrections in backgrounds with RR fields turned on.
Therefore it might be useful to examine theories
where corrections enter at order $1/N$,
which is the topic of this paper (related discussions have
appeared in \cite{Anselmi:1999,Aharony:1999,Blau:1999,Bilal:1999}).

One such case was analyzed in ref.~\cite{Aharony:1999}.
In that paper, it was observed that for the $\cN=2$ superconformal
field theory
with gauge group $\Sp(2N)$ and matter hypermultiplets in the antisymmetric and
fundamental representations,
the $\UR$ anomaly has a subleading $1/N$ correction.\footnote{The
leading order term vanishes \cite{Henningson:1998,Gubser:1998b}.}
On the string theory side, the corresponding anomalous term in the action
was shown to arise from the O7-plane and D7-branes present in the model.

In this paper we extend this result
to a wide class of $\cN=2$ and $\cN=1$ models.
In each case, we find that the $1/N$ correction to the anomaly arises
from the D7-branes and O7-planes present in the dual string theory.
We also consider models without D7-branes and O7-planes,
for which the $1/N$ correction is shown to be absent,
as is required for the consistency of the proposal.

In section \ref{N=2} we describe several classes of type IIB backgrounds,
obtained as orientifolds of $\Z_k$ orbifolds,
and the associated $\cN=2$ superconformal field theories
arising on D3-branes in these backgrounds.
In the large-$N$ limit,
these field theories are dual to type IIB string theory
on $\mathit{AdS}_5{\times} S^5/G$,
where $G$ is the orientifold group.
We calculate the anomaly for the R-current from both the field-
and string-theoretic perspectives, and find agreement at order $1/N$.

In section \ref{N=1} we carry out a similar analysis for IIB string theory
on $\mathit{AdS}_5{\times} T^{11}/\Z_2$,
where $T^{11}=\left [\SU(2){\times} \SU(2)\right]/ \U(1)$
and $\Z_2$ denotes an orientifold group.
We propose a particular orientifold of the conifold,
with an O7-plane and
 D7-branes; D3-branes at the conifold singularity
gives rise to an $\cN=1$ $\Sp(2N){\times}\Sp(2N)$
superconformal field theory.
The anomaly of the R-current is calculated,
with agreement between the field- and string-theoretic descriptions
at order $1/N$.
We also examine a closely related
$\cN=1$ $\Sp(2N){\times}\SO(2N{+}2)$ model,
for which D7-branes are absent.
The $1/N$ correction to the anomaly vanishes, as required for consistency.

In section \ref{summ} we summarize our findings, and
two appendices are devoted to technical matters.

\setcounter{equation}{0}
\section{$\cN=2$ orientifolded orbifold models} \label{N=2}

\subsection{Description of the models}
The $d=4$ $\cN=2$ superconformal field theories we will be studying arise
as the theories on a stack of D3-branes in certain type IIB
orientifold backgrounds.
These field theories are dual, via the AdS/CFT correspondence,
to string theory in the background $\mathit{AdS}_{5}{\times} S^5/G$,
where $G$ is the orientifold group (see \cite{Aharony:1999b} and references therein).
The D3-brane field theories have been constructed in \cite{Park:1998}
and we will use the notations of that paper.
In the T-dual type IIA picture \cite{Witten:1997}, these
models arise from the field theories on D4-branes in elliptic
models  that contain two O6-planes and $k$ NS5-branes.

The field theories fall into two categories.
The first type arises, on the type IIA side,
from the D4-branes in elliptic models with
two O$6^-$-planes, $k$ NS5-branes,
and in addition 4 D6-branes plus their mirrors,
parallel to the orientifold planes.
Via T-duality the D4-branes turn into D3-branes,
the D6-branes turn into D7-branes, the O$6^{-}$-planes
turn into an O7-plane,
and the $k$ NS5-branes dualize into a $\Z_{k}$ orbifold singularity.
The D3-branes span the directions $0123$
and we will label the $456789$ directions transverse
to the D3-branes by three complex coordinates, $(z_3,z_1,z_2)$.
The orientifold group for these theories
has the form $G = G_{\mathrm{orb}}{\times} G_{\mathrm{ori}} $.
The pure orientifold part is $G_{\mathrm{ori}} = \{1,\Om'\}$,
with $\Om' = \Om(-1)^{F_L}R_{45}$, where $R_{45}$ reflects the 45 directions,
\ie{} acts as $z_3 \rar -z_3$ (which implies that the O7-plane
and the D7-branes are located at $z_3=0$).
The orbifold part of the group is
$G_{\mathrm{orb}} = \{1,\tha,\ldots,\tha^{k-1}\}$,
where $\tha$ acts on the 6789 directions as
$(z_1,z_2) \rar (e^{2\pi i/k}z_1,e^{-2\pi i/k}z_2)$.
The gauge groups
of the field theories may be determined
either from the placement of the NS5-branes on the IIA side,
or via the orientifold projections acting on the D3-branes on
the IIB side, and fall into three classes\footnote{In
our conventions the defining  representation
of $\Sp(v)$ is $v$-dimensional; in
particular $\Sp(2)\cong \SU(2)$.} \cite{Park:1998}
\be \ba{lll} \label{--}
(\mathrm{I.i})\quad &
\Sp(v_0)\times \SU(v_1) \times \cdots \times \SU(v_P)\,,
 & k = 2P+1\,, \non\\
(\mathrm{I.ii})&
\Sp(v_0)\times \SU(v_1) \times \cdots \times \SU(v_{P-1})
\times \Sp(v_P)\,, \quad
& k  = 2P\,, \non\\
(\mathrm{I.iii})&
\SU(v_1) \times \cdots \times \SU(v_{P})\,,
& k = 2P \,, \ea \ee
and the matter content is (in $\cN=2$ language)
\bea
&& (\mathrm{I.i}) ~~ \quad \bigoplus_{j=1}^{P} (\Yfund_{j-1},
\Yfund_{j}) + \Yasymm_{P} + \half w_0 \Yfund_0 +
\bigoplus_{j=1}^{P} w_j \Yfund_{j}\,, \non \\
&&
(\mathrm{I.ii}) ~~ \quad \bigoplus_{j=1}^{P} (\Yfund_{j-1},\Yfund_{j})  +
\half w_0\Yfund_0 + \bigoplus_{j=1}^{P-1} w_j \Yfund_{j}
+ \half w_P \Yfund_P \,, \non \\ &&
(\mathrm{I.iii}) ~~ \quad \Yasymm_{1} +\bigoplus_{j=2}^{P}
(\Yfund_{j-1},\Yfund_{j})
+\Yasymm_P + \bigoplus_{j=1}^{P} w_j \Yfund_{j} \,.
\eea
In the above formul\ae{} the $w_j$'s are non-negative integers
related to the placement of the D6-branes, and are
constrained by the equation $\sum_j w_j = 8$,
where the index runs over $k$ values
(although the indexing varies among the different cases).
The $v_j$ are related to the numbers of D4/D3-branes,
and obey constraints (see appendix \ref{Gravanom})
arising from the
vanishing of the beta-function(s) of the field theory.
The AdS/CFT correspondence involves the large $N$ limit,
where $N$ is the total number of D3-branes.
In appendix \ref{Gravanom} we show that, to leading order in $N$,
the $v_j$'s corresponding to each of the group factors
are equal.
We will denote the common value by $v$, which is of order $N$.

The second type of model involves, on the type IIA side,
one O$6^+$ orientifold plane and one O$6^-$ plane,
but no D6-branes,
which implies  the absence of
D7-branes in the corresponding type IIB models.
The form of the orientifold group for these models
can be found in \cite{Park:1998}; it will not be needed for our purposes
(for these models there are no orientifold planes associated with
the orientifold group).
There are four classes of such models, with gauge groups \cite{Park:1998}
\be \ba{lll} \label{+-}
(\mathrm{II.i}) \quad &
\Sp(v_0)\times \SU(v_1) \times \cdots \times \SU(v_P)\,, & k = 2P+1 \,,
\non \\ (\mathrm{II.i^\prime}) &
\SO(v_0)\times \SU(v_1) \times \cdots \times \SU(v_P) \,, & k = 2P+1 \,,
\non \\
(\mathrm{II.ii})&
\SO(v_0)\times \SU(v_1) \times \cdots \times \SU(v_{P-1}) \times \Sp(v_P)
\,, \quad
&k = 2P \,,
\non\\
(\mathrm{II.iii})&
\SU(v_1)\times \cdots \times \SU(v_P) \,, &
 k = 2P \,,
\ea \ee
and $\cN=2$ matter content
\bea
(\mathrm{II.i}) \quad ~~ \bigoplus_{j=1}^{P}
(\Yfund_{j-1},\Yfund_{j}) + \Ysymm_{P} \,, \qquad
&&(\mathrm{II.i^\prime}) \quad ~~ \bigoplus_{j=1}^{P}
(\Yfund_{j-1},\Yfund_{j}) + \Yasymm_{P} \,, \non\\
(\mathrm{II.ii}) \quad ~~ \bigoplus_{j=1}^{P} (\Yfund_{j-1},\Yfund_{j}) \,,
\quad\qquad\qquad
&&(\mathrm{II.iii}) \quad ~~ \Yasymm_{1} +
\bigoplus_{j=2}^{P} (\Yfund_{j-1},\Yfund_{j}) + \Ysymm_{P} \,.
\eea
Again, the $v_j$'s in each theory are equal to leading order in the
large $N$ limit.

All the above field theories have the symmetry group
$\SU(2)_L\times\SU(2)_R\times \UR$,
which is inherited from the $\SU(4)$ R-symmetry group of the $\cN=4$ theory.
The first type of model also has a flavor symmetry arising from the D7-branes;
the flavor symmetry group (which in general is not a simple group)
is a subgroup of $\SO(8)$ for all the models we study.

As mentioned before, the $\cN=2$ field theories described above are
dual to type IIB string theory on $\mathit{AdS}_5{\times} S^5/G$.
On the string theory side the symmetry group arises from the isometry
group of $S^5/G$.
The breaking pattern of $\SO(6)$, the isometry group of $S^5$,
induced by $G$ is
\be \label{branch}
\SO(6) \rar \SO(4){\times} \U(1) \cong \SU(2) {\times}
\SU(2) {\times} \U(1) \,.
\ee
In this equation we have suppressed the action of the
discrete groups on the factors on the right hand side.
The $\U(1)$ factor above is identified with the $\UR$ symmetry group
of the $\cN=2$ field theory, and will play a central role in what follows.

\subsection{Field theory analysis}

We are interested in the
anomaly of the $\UR$ current,
which can be calculated on the field theory side
using standard methods.
The assignments of $\cR$-charges are
as follows\footnote{
These values appear naturally from the branching
rule (\ref{branch}).
}:
in an $\cN=2$ vector multiplet,
the scalar field has $\cR$-charge $q=2$,
whereas the two spin-$\half$ Weyl fermions
have $\cR$-charge $q = 1$;
in an $\cN=2$ hypermultiplet,
the scalar field has $\cR$-charge $q=0 $
and the two spin-$\frac{1}{2}$ fields
both have $\cR$-charge $q = -1$.
The $\cR$-symmetry current is
\be
\label{R}
\cR^{\mu} =
\sum_i \left[
        \half q_i \bar{\psi}_i \ga^\mu \ga_5 \psi_i
       - i q_i \bar{\phi}_i {\mathop{D}^{\leftrightarrow}}{}^{\mu} \phi_i
	\right] \,.
\ee
The anomaly
$\langle \pa_{\mu} ( \sqrt{g} \cR^{\mu})\rangle $
{gets} a contribution from both the coupling to gravity and
to the flavor group.

For the gravitational part,
a single Weyl fermion with $\cR$-charge $q$ contributes
$q R \tilde{R}/384\pi^2 $
to the anomaly \cite{Alvarez-Gaume:1984},
where
$R \tilde{R} \equiv \half
\epsilon_{\mu\nu\alpha\beta}
{R^{\mu\nu}}_{\sigma\tau}
R^{\alpha\beta\sigma\tau}$.
Each vector multiplet $V_i$ contains two Weyl fermions of $\cR$-charge
$+1$ and each hypermultiplet $H_j$ contains two Weyl fermions
of $\cR$-charge $-1$
so the total gravitational anomaly is
\be
\label{Qdef}
\langle \pa_{\mu} ( \sqrt{g} \cR^{\mu})\rangle_{\rm grav}
=  {Q\over 384\pi^2} R \tilde{R}
= \frac{2}{384\pi^2}(\sum_i {\rm dim}~ V_i -  \sum_j {\rm dim}~ H_j)
R \tilde{R} \,.
\ee
In appendix A,  we explicitly compute $Q$ for each of the
theories described in sec.~2.1.
In the large $N$ limit, the gravitational anomaly is at first sight
 $\cO(N^2)$, but the leading order contribution cancels
for all the models we study in this paper.
The remaining $\cO(N)$ contribution for each of the
models (I.i) through (I.iii) is given by the universal expression
\be
\langle \pa_{\mu} ( \sqrt{g} \cR^{\mu})\rangle_{\rm grav}
= -\frac{v}{64\pi^2} R \tilde{R}\,,
\ee
where $v$ was defined in sec.~2.1.
The $\cO(1)$ contribution to the anomaly differs among these models,
but will not be relevant for our purposes.

There is also a contribution to the anomaly
induced from the triangle diagram constructed from the $\cR$-current
and  two flavor currents.
The flavor current is
\be \label{J}
J_{\mu}^a
= \sum_{i} \left[-{\ts \frac{1}{2} } \bar{\psi}_i \ga_{\mu}(1-\ga^5)t^a \psi_i
+ \bar{\phi}_i\mathop{D}^{\leftrightarrow}{}_{\mu} t^a \phi_i\right]\,,
\ee
where the sum runs over those hypermultiplets
carrying flavor quantum numbers.
As the number of hypermultiplets
for each factor of the flavor group is $v$ (to leading order),
one finds the anomaly
\be
\langle \pa_{\mu} ( \sqrt{g} \cR^{\mu})\rangle_{\rm flavor}
= \frac{v}{16 \pi^2} F \tilde{F} \,,
\ee
where
$F \tilde{F} \equiv \half  \epsilon_{\mu\nu\kappa\lambda}
\tr( F^{\mu\nu} F^{\kappa\lambda})$,
and where $\tr$ denotes the trace
over the representation of the total flavor group
into which SO(8) branches.
Hence, the total $\UR$ anomaly for models (I.i) through (I.iii),
correct to $\cO(N)$, is
\be
\label{totalanom}
\langle \pa_{\mu} ( \sqrt{g} \cR^{\mu})\rangle_{\rm total}
= -\frac{v}{64\pi^2} (R \tilde{R} - 4 F \tilde{F})\,.
\ee
For the $k=1$ model of type (I.i), eq.~(\ref{totalanom})
agrees with the result given in ref. \cite{Aharony:1999}.

In the second class of models, (II.i) through (II.iii),
the gravitational contribution to the  $\UR$ anomaly
is shown in appendix A to vanish at $\cO(N)$,
and there is no flavor group to contribute to the anomaly.

We will argue, following \cite{Aharony:1999},
that in the dual string theory
the $\cO(N)$ contributions to the $\UR$ anomaly
arise solely from the presence of
O7-planes and  D7-branes in the models.
This proposal is consistent with the absence of
$\cO(N)$ contributions to the anomaly for the second class of models
(\ref{+-}), since they have no D7-branes
or orientifold planes.

\subsection{Supergravity analysis}
We now calculate the $\cR$-current anomaly
from the dual supergravity theory,
closely following the analysis of ref.~\cite{Aharony:1999}.
We will need the $R^2$ and $F^2$ couplings in the
Wess-Zumino (a.k.a. Chern-Simons) term in the world-volume action
for the D7-branes.
These couplings are obtained from the
result\footnote{There is a well known ordering ambiguity for the
non-abelian gauge fields; fortunately, for the terms we need this
problem does not arise.} \cite{Douglas:1995}
\be
S_{WZ} = \mu_7
\int C \we \sqrt{\hat{A}(4\pi^2\al'R)}\tr\, {\rm e}^{2\pi\al'F} \Big|_8 \,,
\ee
where $\mu_7 =({2(2\pi)^7\al'^4})^{-1} $  is the
charge of a single D7-brane \cite{Aharony:1999},
$C=\sum C_{n}$ is the sum over antisymmetric RR form fields,
$F$ is the
field strength of the non-abelian gauge field on the D7-branes,
and the integration picks out the eight form.
There is also a contribution from the  O7
orientifold planes which are coincident with the
D7-branes \cite{Dasgupta:1998}
\be
S_{O} = \mu_7^\prime \int C \we \sqrt{\hat{L}(\pi^2\al'R)} \Big|_8 \,,
\ee
where $\mu_7^\prime = -2^3 \mu_7$  \cite{Aharony:1999}.
For the eight D7-branes and one O7-plane present in models
(I.i) through (I.iii),
one extracts from the above expressions
the terms relevant to the anomaly \cite{Aharony:1999}
\be \label{WZact}
\frac{1}{2^9 \pi^5 \al'^2}\int C_4 \we [\tr (R\we R) +  4\,\tr (F\we F)]\,,
\ee
using the fact that $\tr(\1) = 8$.

Next, we need to rescale the $C_4$ field to agree
with AdS/CFT conventions.
In the usual
supergravity convention, the relevant part of the
type IIB supergravity action is
\be
{1\over 2 \ka^2} \int \D^{10} x \sqrt{-g}
\left[R - \frac{1}{4\cdot5!}g^2_{\mathrm{s}}F_{5}^{2} \right] \,.
\ee
The rescaling
$g_{\mu\nu} \rar L^2 g_{\mu \nu}$
together with
$C_4 \rar 4 L^4 g_{\mathrm{s}}^{-1} C_4$
puts the action into the usual AdS/CFT form
\be
\label{lengthscale}
{L^{8} \over 2\ka^2} \int \D^{10} x \sqrt{-g}
\left[R - \frac{1}{5 \cdot 3!}F_{5}^{2}\right].
\ee
Here $L$ is the fundamental length scale defined by \cite{Gubser:1998b}
\be
\label{Ldef}
L^4 = \frac{\sqrt{\pi} \ka N_{\mathrm{phys}}}{2 \mathrm{Vol}(X^5) } \,,
\ee
where $X^5 = S^5/G$ and $N_{\mathrm{phys}}$ is the number of physical branes.
Since $\mathrm{Vol}(X^5) = \mathrm{Vol}(S^5)/2k$,
and $N_{\mathrm{phys}}= N_{\mathrm{cover}}/2k $,
where $N_{\mathrm{cover}}$ is the number of branes on the cover space,
we may rewrite this as
\be
L^4 = \frac{\sqrt{\pi} \ka N_{\mathrm{cover}}}{2 \mathrm{Vol}(S^5) }\,.
\ee
In other words, $L^4$ is the same before and after the projection.
The sum over branes and their mirrors
$N_{\mathrm{cover}} = \sum_j v_j$
is equal to  $kv$ to leading order in $N$,
so to leading order in $N$, we have
\be
\label{Lleading}
L^4 = \frac{\ka k v} {2 {\pi}^{5/2}}\,,
\ee
where $\mathrm{Vol}(S^5)  = \pi^3$.
Using $2\ka^2 = (2\pi)^7 g_{\mathrm{s}}^2\al'^4$,
the rescaling becomes
$C_4 \rar  16\pi \al'^2 k vC_4$,
thus changing (\ref{WZact}) to
\be
\label{WZactrescale}
\frac{kv}{32 \pi^4 }\int C_4 \we [\tr (R\we R) +  4\,\tr (F\we F)]\,.
\ee
The world volume of the D7-branes is $\mathit{AdS}_5{\times} S_3/\Z_k$,
where $S^3/\Z_k$ is
a completely non-singular Lens space;
the action of $\Z_k$ on the 6789 coordinates was discussed
in sec.~2.1.

We will now reduce the action (\ref{WZactrescale})
to five dimensions;
since we are interested in the massless vector
corresponding to the $\UR$ symmetry
we will only keep this mode when reducing.
Massless vectors arise from the $g_{\mu a}$ and $C_{\mu abc}$
components of the metric and four-form \cite{Kim:1985}.
In the expansion of these fields in vector harmonics,
only the first level ($j=1$) contributes;
we denote the harmonic at this level appropriate to the
$\UR$ gauge field as $Y_{a}$
and write
\bea
\label{masslessmodes}
g_{\mu a} &=& B_{\mu}Y_{a} \,, \non \\
C_{\mu abc} &=& \phi_{\mu} \sqrt{G} \ep_{abc}{}^{de}D_{d}Y_{e} \,.
\eea
Using the explicit form of $Y_a$ in ref.~\cite{Aharony:1999}
and restricting $C_{\mu abc}$ to the D7-brane world-volume,
we find
\be
\label{C}
C_{\mu abc} = \phi_{\mu} \omega_{abc} \,,
\ee
where $\om_{abc}$ is
the volume form on $S^{3}/{\Z_k}$.

It was shown in \cite{Kim:1985} that
the linear combination $A_{\mu}= B_{\mu}-16\phi_{\mu}$ is massless.
This is the mode that corresponds to the $\UR$ vector field,
up to a normalization factor,
\be
\label{ARdef}
A_\mu = -24 \eta  A^{\cR}_{\mu} \,,
\ee
where $A^{\cR}_{\mu}$  is the $\UR$ gauge field canonically coupled
to the $\cR$-current (\ref{R}),
and $\eta$ remains to be determined.
(The $\UR$ gauge field is present in all the models we consider,
as the orbifold generator does not act on this mode.)
Setting the orthogonal combination $V_{\mu} = B_{\mu} + 8\phi_{\mu}$
(which is massive)  to zero, we have
\be
\label{eta}
\phi_\mu = \eta A^{\cR}_{\mu}.
\ee
Using  eqs.~(\ref{C}) and (\ref{eta}) in (\ref{WZactrescale}),
and integrating over $S^3/{\Z_k}$
 ($\mathrm{Vol}(S^3/\Z_k) = {2\pi^2}/{k}$),
we obtain
\be \label{5dact}
\eta \frac{v}{16\pi^2}\int_{{\rm AdS}_5} \!\!
A^{\cR} \we [\tr (R\we R) +  4\, \tr (F\we F)]\,;
\ee
note that the factors of $k$ have cancelled.
This expression is not invariant under a $\UR$ gauge transformation
$A^{\cR} \rar A^{\cR} + \D \La$,
but gives a boundary contribution
\be
\label{4dact}
-\eta \frac{v}{16\pi^2}
\int_{M_{3,1}} \!\!\!\!
\La  [\tr (R\we R) +  4\, \tr (F\we F)]
=
\eta \frac{v}{32\pi^2}
\int_{M_{3,1}} \!\!\!\! \D^4 x
\sqrt{-g} \,
\La  (R \tilde{R}  -  4\, F \tilde{F} )\,.
\ee
This implies that the anomaly is
\be
\label{stringanom}
\langle \pa_{\mu} ( \sqrt{g} \cR^{\mu})\rangle_{\rm total}
= \,-\,\eta \frac{v}{32\pi^2}  (R \tilde{R}  -  4\, F \tilde{F} )\,.
\ee
We will now show that this result agrees with the field-theoretic
result (\ref{totalanom}).

First, we must verify that the relative factor between
the gravitational and flavor contributions agrees for the
field and string theory calculations.
This appears to be the case
from comparison of the formul\ae{}  (\ref{totalanom}) and (\ref{stringanom}).
However, as was stressed in \cite{Aharony:1999},
one needs to check that the generators of the flavor symmetry group
have the same normalization in both descriptions.
We will apply the line of reasoning in
\cite{Aharony:1999} to each factor of the flavor group.
We begin by expanding the Born-Infeld action for the D7-branes
to second order in the gauge-field:
\be
\label{BI}
S_{\mathrm{BI}}
= -\mu_7 \int \D^{8}x\, e^{-\Phi}\
\tr \sqrt{-\det(g+2\pi\al'F)} ~~\rar ~~
 \mu_7 (2\pi\al')^2 \int \D^{8}x \sqrt{-g}\, e^{-\Phi}\,{1\over 4}\,\tr F^2 \,.
\ee
Let the Lie algebra generators $T^a$ appearing in
eqs.~(\ref{5dact}) and (\ref{BI}) be normalized as
$\tr(T^a T^b) = \la \de^{ab}$.
Rescaling the metric $g\rar L^2 g$  in eq.~(\ref{BI})
and dimensionally reducing on $S^3/\Z_k$ yields
\be \label{5dYMact}
\la \mu_7 e^{-\Phi}(2\pi\al')^2 L^4 \mathrm{Vol}(S^3/\Z_k)
\int \D^{5}x \frac{1}{4}\sqrt{-g}\, F^a_{\mu\nu}F^{a\mu\nu}
\equiv
\frac{1}{4g_{\mathrm{F}}^2}
\int \D^{5}x \sqrt{-g}\, F^a_{\mu\nu}F^{a\mu\nu} \,.
\ee
Using $\mu_7 e^{-\Phi} = \frac{ \sqrt{\pi} }{2\ka} ( 4\pi^2\al')^{-2}$
and
$\mathrm{Vol}(S^3/\Z_k) = {2\pi^2}/{k}$,
together with eq.~(\ref{Lleading}),
we deduce that,
for each subgroup of the flavor group, the
associated flavor symmetry coupling
constant
${1}/{g^2_{\mathrm{F}}}$  is equal to
${\la v}/{2(2\pi)^2}$.

Next, following \cite{Aharony:1999},
we determine the normalization of the generators $t^a$ appearing in
the field theory current (\ref{J}).
The expression for the correlator of two flavor currents (\ref{J})
predicted by the AdS/CFT correspondence is given by \cite{Freedman:1998}
\be \label{JJ1}
\langle J^a_{\mu}(x) J^{b}_{\nu}(0) \rangle =
{\de^{ab} \over 2\pi^2 g^2_{\mathrm{F}}}
(\eta_{\mu\nu} \Yfund -  \pa_{\mu}\pa_{\nu})
\frac{1}{x^4}
=
{v \over (2\pi)^4}
\la  \de^{ab} (\eta_{\mu\nu} \Yfund -  \pa_{\mu}\pa_{\nu})
\frac{1}{x^4}\,.
\ee
The correlator of two currents (\ref{J}) calculated directly in field theory
is
\be \label{JJ2}
\langle J^a_{\mu}(x) J^{b}_{\nu}(0) \rangle =
{v \over (2\pi)^4 }
\tr(t^a t^b)
(\eta_{\mu\nu} \Yfund -  \pa_{\mu}\pa_{\nu})
\frac{1}{x^4}\,,
\ee
using the fact that for each factor of the flavor group
the number of hypermultiplets is $v$.
Comparing eqs.~(\ref{JJ1}) and (\ref{JJ2}),
we find that $\tr(t^a t^b) =  \la \de^{ab}$.
That is, the field theory normalization of generators
is the same as the string theory normalization, and
so the relative coefficients of the $\UR$ anomaly are the
same on both sides of the AdS/CFT correspondence.

Next, we will verify that the overall normalization of the anomaly
agrees for the field and string theory calculations.
Following the approach in \cite{Aharony:1999},
we begin with the action for the massless vector
$A_{\mu}$ obtained from the KK reduction of the type IIB supergravity action,
found in \cite{Arutyunov:1998},
\be \label{5dredact}
S = \frac{L^8}{2\ka^2}
\int_{X_5} \!\!\!\!\D^5y \sqrt{G} \,G_{ab} Y^aY^b
\int_{\mathit{AdS}_5}\!\!\!\! \D^5 x \sqrt{-g}
\frac{1}{3}\left[- \frac{1}{4}  F(A)^2  \right]
\equiv -\frac{1}{4g_{\mathrm{SG}}^2}\int \D^5x \sqrt{-g} F(A^{\cR})^2\,,
\ee
where we have rescaled the ten-dimensional metric as $g\rar L^2 g$.
The factor $\frac{1}{3}$ arises from the $\half (j+1)/(j+2)$ factor
in eq.~(3.17) of ref.~\cite{Arutyunov:1998} with $j=1$,
which corresponds to the massless field.
Using eqs.~(\ref{Lleading}) and (\ref{ARdef})
together with
$\int \D^5y \sqrt{G}\, G_{ab} Y^a Y^b = {\pi^3}/{24k}$,
we obtain
$ {1}/{g^2_{\mathrm{SG}}} = {\eta^2 k v^2}/{\pi^2}$.

Next we will calculate the 2-point function of the $\cR$-current in two ways.
First, it can be obtained using the AdS/CFT
correspondence \cite{Freedman:1998}
\be
\label{RR1}
\langle \cR_\mu(x)\cR_\nu(0) \rangle =
\,-\,\frac{1}{2\pi^2 g_{\mathrm{SG}}^2}
(\eta_{\mu\nu} \Yfund -  \pa_{\mu}\pa_{\nu})
\frac{1}{x^4}
=
\,-\, \frac{\eta^2 k v^2}{2\pi^4}
(\eta_{\mu\nu} \Yfund -  \pa_{\mu}\pa_{\nu})
\frac{1}{x^4}\,.
\ee
Second, it can be obtained from a purely field-theoretic calculation
(\eg{} using the methods in \cite{Anselmi:1997})
with the result\footnote{Note
that the R-charge assignments in \cite{Anselmi:1997}
are those appropriate for an $\cN=1$ theory.}
\be
\label{RR2}
\langle \cR_\mu(x)\cR_\nu(0) \rangle = \,-\, \frac{2k v^2}{(2\pi)^4}
(\eta_{\mu\nu} \Yfund -  \pa_{\mu}\pa_{\nu})
\frac{1}{x^4}\,.
\ee
Comparing eqs.~(\ref{RR1}) and (\ref{RR2})
we find $\eta = \frac{1}{2}$,
which is precisely the value needed for agreement of
the field theory and string theory calculations of the anomaly.

\setcounter{equation}{0}
\section{$\cN=1$ orientifolded conifold models} \label{N=1}

\subsection{Description of the models}

All models described in section 2 involve string theory on
$\mathit{AdS}_5{\times} X_5$, where $X_5$ is locally $S^5$.
The simplest known example of a dual pair where $X_5$ is not locally $S^5$ is
the duality between the field theory on D3-branes at a
conifold singularity and
type IIB string theory on $\mathit{AdS}_5{\times} T^{11}$
\cite{Klebanov:1998}.

The conifold can be described by the equation $xy=wz$ in \,$\C^4$.
(Another
common way to write the equation is $z_1^2 + z_2^2 + z_3^2 + z_4^2 = 0$;
the two equations are related by a simple linear change of basis.)
This equation defines a cone with a singularity at the apex.
The base of the cone is obtained by intersecting the above
space with
$|x|^2 + |y|^2 + |z|^2 + |w|^2 = 1$,
and can be shown to be $T^{11}$ \cite{Candelas:1990},
which is the coset space $[{\SU(2)\times \SU(2)}]/{\U(1)}$.
This coset space is topologically (but not metrically) $S^3{\times} S^2$
and locally $S^2{\times} S^2 {\times} S^1$.
A useful parameterization of $T^{11}$ is
\bea \label{xyzw}
&&
x = \cos {\half \thaa } \cos {\ts \half \thaap }
\e^{\frac{i}{2}(\psi + \phii + \phiip)}\,, \qquad
y = \sin {\ts \half \thaa } \sin {\ts \half \thaap }
\e^{\frac{i}{2}(\psi - \phii - \phiip)}\,, \non \\ &&
w = \cos {\ts \half \thaa }\sin {\ts \half \thaap }
\e^{\frac{i}{2}(\psi + \phii - \phiip)}\,, \qquad
z = \sin {\ts \half \thaa }\cos {\ts \half \thaap }
\e^{\frac{i}{2} (\psi - \phii + \phiip)} \,,
\eea
with the metric \cite{Candelas:1990}
\be
\label{T11met}
\D s^2 = \ninth (\D \psi + \cos\thaa\D \phii + \cos \thaap \D \phiip)^2
+ \sixth(\D \thaa^2 + \sin^2\thaa \D\phii^2)
+ \sixth(\D \thaap^2 + \sin^2\thaap \D\phiip^2)\,,
\ee
where the range of the variables is $\thaa$, $\thaap \in [0,\pi)$,
$\phii$, $\phiip \in [0,2\pi)$
and $\psi\in[0,4\pi)$.
This metric has determinant
\be
\label{T11det}
\sqrt{G} =  {\ts {1\over 108}} \sin\thaa \sin\thaap\,,
\ee
yielding $\mathrm{Vol}(T^{11}) = {16\pi^3}/{27}$.

The $N$ D3-branes at the conifold singularity
give rise to an $\cN=1$ conformal field theory,
with gauge group $\SU(N){\times} \SU(N)$,
matter chiral multiplets in the representations
$2 (\Yfund,\overline{\Yfund}) \oplus 2(\overline{\Yfund},\Yfund)$,
and symmetry group $\SU(2){\times} \SU(2) {\times} U(1)_{R}$,
where $\U(1)_R$ is the R-symmetry group of the $\cN=1$ field theory.

As shown in \cite{Klebanov:1998}
the conifold model can be obtained by
a perturbation of the model corresponding to
$N$ D3-branes at a $\Z_2$ orbifold singularity,
which is dual \cite{Kachru:1998} to type IIB
string theory on $\mathit{AdS}_5{\times} S^5/\Z_2$.
On the string theory side,
the perturbation is realized as a blow up of
$S^5/\Z_2$ to the smooth space $T^{11}$.
On the field theory side, the perturbation corresponds to a
renormalization group flow from the $\Z_2$ orbifold field theory in the UV
to the conifold field theory in the IR.
The $\Z_2$ orbifold field theory is an $\cN=2$ conformal field theory
with gauge group $\SU(N){\times} \SU(N)$,
whose matter content (in $\cN=1$ language) consists of
chiral multiplets in the representations
$({\rm adj}, 1) \oplus (1, {\rm adj})
\oplus 2 (\Yfund,\overline{\Yfund}) \oplus 2(\overline{\Yfund},\Yfund)$.
The RG flow can be realized by integrating out the adjoint matter fields.
It has been shown \cite{Gubser:1998b} that the flow of the
leading order ($\cO(N^2)$) correction to the conformal anomaly
can be explained from the gravity side
via the fact that the central charge
is inversely proportional to the volume of $X_5$.

As in section 2, we would like to study ${1}/{N}$
corrections to the $\U(1)_R$ anomaly.
The above model has no such corrections;
to obtain models that do,
we will consider orientifolds of the conifold model.

We start by describing the introduction of orientifold planes
on the type IIA side,
as that provides a more intuitive understanding.
The field theory corresponding to the conifold singularity
arises on the IIA side from an elliptic model
with two NS5-branes rotated 90 degrees with respect to
one another \cite{Brunner:1998}.
Such a configuration can be obtained from the IIA model with two parallel
NS5-branes (which corresponds to the type IIB $\Z_2$ orbifold model)
by rotating the NS5-branes.
This is the type IIA analog of
perturbation of the $\Z_2$ orbifold model
to obtain the conifold model.

One can introduce orientifold (O6) planes into the
IIA analog of the type IIB $\Z_2$ orbifold.
There are four such models, which are described in sec.~2.1
(see \eg{} \cite{Ennes:2000} for a more detailed discussion).
Two of these models correspond to placing the O6-planes
away from the two NS5 branes along the compact direction
and the other two correspond to placing
one NS5-brane on top of each O6-plane.
The first two models can be related to orientifolded
conifold models \cite{Brunner:1998} by rotating
the NS5-branes 45 degrees in opposite directions
(so that the angle between them becomes 90 degrees)
while respecting the orientifold reflection\footnote{This is not possible
for the other two models since the rotation is
not compatible with the orientifold reflection.
The configuration with one of the two NS5-branes tilted 90 degrees
is allowed, but is not continuously connected to
the orbifold model via a rotation. }.
The resulting gauge theories have the following gauge groups and
$\cN=1$ matter content
\be \label{matter} \ba{lll}
\mbox{(i)} & \Sp(v){\times} \Sp(v)\,, & \mbox{with matter}
\quad2(\Yfund,\Yfund)\oplus 4(\Yfund,1) \oplus 4(1,\Yfund)\,, \\[1mm]
\mbox{(ii)} &\Sp(v){\times}\SO(v+2)\,,& \mbox{with matter}
\quad 2(\Yfund,\Yfund)\,.
\ea \ee
Since the first model has eight D6-branes parallel to the orientifold planes,
the type IIB version of this model is a possible candidate for a
theory with D7-branes and $1/N$ corrections to the $\U(1)_R$ anomaly.
However, in general it is not known how to T-dualize type IIA models
based on non-orbifold singularities\footnote{Another approach
based on blowing up orbifold singularities was pursued in \cite{Uranga:1999},
where some type IIB orientifolds of the conifold were constructed,
though not the particular models we are interested in.
Although it should be possible to generalize the analysis of
\cite{Uranga:1999}
to incorporate the models we study in this paper,
we will not attempt to do so here. }.

Instead,  we will assume that the orientifolded conifold model
can be obtained by a perturbation of the model corresponding to
D3-branes at an orientifolded $\Z_2$ orbifold singularity.
The orientifolded $\Z_2$ orbifold field theory is an
$\cN=2$ conformal field theory of type (I.ii) from sec. 2.1,
with gauge group $\Sp(v) {\times} \Sp(v)$,
and whose matter content (in $\cN=1$ language) consists of
chiral multiplets in the representations
$({\rm adj}, 1) \oplus (1, {\rm adj}) \oplus 2
(\Yfund,\Yfund)\oplus4(\Yfund,1)\oplus 4(1,\Yfund) $.
We further assume that, on the field theory side,
the perturbation corresponds to a
renormalization group flow from the orientifolded $\Z_2$
orbifold field theory in the UV
to the orientifolded conifold field theory with field
content (\ref{matter}i) in the IR.
It seems likely that this will work in an analogous way
to the unorientifolded case, with the RG flow
realized by integrating out the adjoint matter fields,
but there are cases
for which the method of integrating out adjoint chiral
superfields gives the correct matter content but an
incorrect superpotential (see \eg{} the disussion in \cite{Uranga:1999}).
Therefore, we will make no assumptions about
the exact form of the superpotential or whether
the RG flow can be realized by integrating out adjoint chiral multiplets.
We will, however,  assume that the anomalous dimensions of
the fields are the same as those that would result from such a procedure.

In the following, we will need
the action of the orientifold on the coordinates of  $T^{11}$.
As noted in \cite{Uranga:1999} it is usually possible to deduce
the orientifold action in the type IIB model from its IIA analog,
without going through the complete construction.
On the type IIA side, the rotation of
the branes is easily incorporated into the equation for
the orbifold\footnote{Recall that the invariant coordinates for
the $\C^2/{\mathsf{Z}\kern -4pt \mathsf{Z}}_2\times \C$ orbifold
are $x$, $y$, $\tilde{w}$ and $\tilde{z}$,
where $xy=\tilde{w}^2$.}
as $x y = (\tilde{w}\cos \al_1 + \tilde{z}\sin\al_1)
(\tilde{w}\cos\al_2 + \tilde{z} \sin\al_2)$.
For the $\Sp(v){\times}\Sp(v)$ orientifolded $\Z_2$-orbifold,
the orientifold acts on the invariant coordinates
as $\tilde{z}\rar -\tilde{z}$,
with the other coordinates unaffected.
This action is compatible with the rotation only if  $\al_1 = - \al_2$.
In particular, for $\al_1 = - \al_2={\pi}/{4} $
we find $x y = wz$ where
$w = \frac{1}{\sqrt{2}}(\tilde{w} + \tilde{z})$
and $z = \frac{1}{\sqrt{2}}(\tilde{w} - \tilde{z})$.
The inherited orientifold action on these variables is $w\leftrightarrow z$,
with $x,y$ invariant.
In terms of the angles in the parameterization (\ref{xyzw}) the action is
$\thaa\leftrightarrow\thaap$ and $\phii\leftrightarrow\phiip$.
The fixed point set of this action is
a three-dimensional subspace of $T^{11}$,
which we will denote $X_3$, defined by the
equations $\thaa=\thaap$, $\phii=\phiip$.
The metric on $X_3$ is
\be \label{X3}
\D s^2 = \ninth(\D \psi + 2\cos \thaa \D \phii)^2
       + \third(\D \thaa^2 + \sin^2\thaa \D\phii^2)\,,
\ee
where the ranges of the variables are $\thaa \in [0,\pi)$,
$\phii \in [0,2\pi)$, and $\psi\in[0,4\pi)$.
The determinant of this metric is
\be
\label{X3det}
\sqrt{G_3} = \ninth \sin\thaa\,,
\ee
and the volume of $X_3$ is $16 \pi^2/9$.

We need not consider the orientifold action of
the $\Sp(v) {\times} \SO(v+2)$ model,
since that model has no D7-branes.

\subsection{Field theory analysis}
In this section we will calculate the anomaly
in the $\Ur$ current in the $\cN=1$ field theories
corresponding to D3-branes at an orientifolded conifold singularity
described in sec.~3.1.
The $\cN=1$ R-charge assignments are as follows:
the fermions in each $\cN=1$ vector multiplet have R-charge $+1$,
whereas the fermions (resp. bosons) in each $\cN=1$ chiral multiplet
have R-charge $-\frac{1}{3}$ (resp. $\frac{2}{3}$).
(The $\cN=2$ $\cR$-charge assignments of sec.~2 are related to the
$\cN=1$ R-charge assignments via $\cR= 3R - 4 J_z$,
where $J_z$ is the $z$-component of the $\cN=2$ $\SU(2)_R$ symmetry.)

Anomalies in orbifold theories can easily
be calculated using methods of sec.~\ref{N=2}.
We are interested in how the anomaly changes under the RG flow
to the IR fixed point.
Although the R-current is not invariant under the RG flow,
the following combination
\be
\label{S}
S_{\mu} = R_{\mu} + \frac{2}{3}\sum_i (\ga^i_{\mathrm{IR}} - \ga^i)K^i_{\mu}
\ee
of the R-current and the Konishi current
$K_{\mu}^i = \frac{1}{2}\bar{\psi}_i\ga_\mu \ga_5 \psi_i
- \bar{\phi}_i { \stackrel{\leftrightarrow}{D}_{\mu} } \phi_i$
is invariant \cite{Anselmi:1998,Gubser:1998b,Aharony:1999b},
where the sum in (\ref{S}) runs over the $\cN=1$ chiral multiplets.
Here $\ga_{\mathrm{IR}} = -\frac{1}{4}$ for the $\cN=1$ chiral superfields
in the $\cN=2$ hypermultiplets and $\ga_\mathrm{IR} = \frac{1}{2}$
for the $\cN=1$ chiral superfield in the $\cN=2$ vector multiplet.
Thus, to calculate the anomalies of the R-current in the IR fixed point
theory, one uses that $S_{\mu} = R_{\mu}$ at the IR fixed point,
and that anomalies involving $S_{\mu}$ are invariant under the RG flow,
so that one can calculate the anomaly of $S_{\mu}$ in the UV fixed point
theory
\be \label{method}
 \langle \pa_{\mu} ( \sqrt{g} R^{\mu})\rangle_{\mathrm{IR}}
=\langle \pa_{\mu} ( \sqrt{g} S^{\mu})\rangle_{\mathrm{IR}}
=\langle \pa_{\mu} ( \sqrt{g} S^{\mu})\rangle_{\mathrm{UV}} .
\ee
The S-charges in the UV (where $\ga=0$) are:
$+1$ for fermions in $\cN=1$ vector multiplets,
$-\frac{1}{2}$  for fermions in $\cN=1$ chiral multiplets
coming from $\cN=2$ hypermultiplets,
and $0$ for fermions in $\cN=1$ chiral multiplets
coming from $\cN=2$ vector multiplets.
Thus, the latter do not contribute to the anomaly triangle diagrams,
as is required for consistency, since they are integrated out.

In the ultraviolet,
the $\Sp(v){\times}\Sp(v)$ model
has $v(v+1)$ vector multiplets,
$2v^2 + 8v$ chiral multiplets coming from $\cN=2$ hypermultiplets,
and $v(v+1)$ chiral multiplets coming from $\cN=2$ vector multiplets.
Using the S-charges above,
we find that the gravitational contribution to the $\Ur$ anomaly
for the $\Sp(v){\times} \Sp(v)$ model is
\be
\langle \pa_{\mu} ( \sqrt{g} R^{\mu})\rangle_{\mathrm{grav}}
= -\frac{3v}{384\pi^2} R\tilde{R} \,.
\ee
Since we have $v$ flavors
in each factor of the $\SO(4){\times}\SO(4)$ flavor group of
the UV theory, the flavor contribution to the anomaly is
\be
\langle \pa_{\mu}(\sqrt{g}R^{\mu})\rangle_{\mathrm{flavor}}
= ({\ts \frac{1}{2} })\frac{v}{16 \pi^2}F\tilde{F} \,,
\ee
where we used the relations (\ref{method}) and the $\frac{1}{2}$ comes from
the change in charges compared to section \ref{N=2}.
The total anomaly is therefore
\be
\label{totalanom2}
\langle \pa_{\mu}(\sqrt{g}R^{\mu})\rangle_{\mathrm{total}}
= -\frac{v}{128\pi^2}( R\tilde{R} -4 F\tilde{F}) \,.
\ee

For the $\Sp(v){\times} \SO(v+2)$ model
we have $v^2+2v+1$ vector multiplets coming from the $\cN=2$
vector multiplets, $2v^2 + 4v$ chiral multiplets coming from $\cN=2$
hypermultiplets and (in the UV) $v^2 + 2v + 1$ chiral multiplets
coming from $\cN=2$ vector multiplets.
Using the above method, one finds that
the $\cO(N)$ correction to the $\Ur$ anomaly vanishes.
This is required for consistency of our picture,
since there are no D7-branes
or orientifold planes present in the dual string theory.

\subsection{Supergravity analysis}
We now turn to the calculation of the anomaly in the supergravity dual.
As in section \ref{N=2} the relevant terms from the Wess-Zumino
terms in the worldvolume
action of the D7-branes and O7-plane are
\be \label{WZact2}
\frac{1}{2^9 \pi^5 \al'^2}\int C_4 \we [\tr (R\we R) +  4\,\tr (F\we F)]\,.
\ee
As before, we rescale the field
$C_4 \rar 4 L^4 g_{\mathrm{s}}^{-1} C_4$,
where $L$ is given by (\ref{Ldef}),
to agree with AdS/CFT conventions.
{}From the discussion in section \ref{N=2},
we recall that $L$ is unchanged
when dividing the space by a discrete group,
so we have \cite{Gubser:1998b}
\be
\label{Lconifold}
L_{T^{11}}^4 = \frac{\sqrt{\pi} \ka v }{2 \mathrm{Vol}(T^{11})}
= \left(\frac{3}{2}\right)^3 \frac{\ka v}{ (2\pi)^2\sqrt{\pi} }\,.
\ee
Using $2\ka^2 = (2\pi)^7 g_{\mathrm{s}}^2\al'^4$,
the required rescaling becomes
$C_4 \rar  3^3 \pi \al'^2 v C_4$,
upon which eq.~(\ref{WZact2}) becomes
\be
\label{WZactrescale2}
\frac{3^3 v}{2^9\pi^4 }\int C_4 \we [\tr (R\we R) +  4\,\tr (F\we F)]\,.
\ee
The world volume of the D7-branes is $\mathit{AdS}_5{\times} X_3$.

To reduce the action (\ref{WZactrescale2}) to five dimensions,
we need to expand $C_4$ in terms of vector harmonics on $T^{11}/\Z_2$.
Since we are interested in the massless vector
corresponding to the $\Ur$ symmetry,
we will only keep this mode when reducing. As in section \ref{N=2}
the massless vector arises
from the fields (\ref{masslessmodes}).
The KK reduction on $T^{11}$ has been worked out in detail
in refs.~\cite{Ceresole:1999a,Gubser:1998b,Jatkar:1999}.
In appendix B, we apply these results to find the vector harmonic
$Y_{a}$ corresponding to the $\Ur$ massless vector field.
Since this vector harmonic is invariant under the $\Z_2$ orientifold
operation it is also present in $T^{11}/\Z_2$ model.
Thus
\bea
C_{\mu abc} &=& \phi_{\mu} \sqrt{G} \ep_{abc}{}^{de}D_{d}Y_{e}  \non\\
&=& \phi_{\mu} \sqrt{G}
\left[ \ep_{abc \thaa\phii} G^{\thaa\thaa} G^{\phii\phii}
(\partial_\thaa Y_\phii - \partial_\phii Y_\thaa)
 + \ep_{abc \thaap\phiip} G^{\thaap\thaap} G^{\phiip\phiip}
(\partial_\thaap Y_\phiip - \partial_\phiip Y_\thaap)\right] \non\\
& = & \ninth \phi_\mu
\left(  \ep_{abc \thaa\phii} \sin \thaa
      + \ep_{abc \thaap\phiip} \sin \thaap \right)\,,
\eea
where we have used the $T^{11}$ metric (\ref{T11met}),
its determinant (\ref{T11det}) and the form of $Y_a$ (\ref{vectorhmc}),
and have only kept the terms that give a non-zero contribution
upon restricting to $X_3$.
Restricting $C_4$ to $X_3$ --- the fixed point set
$\thaa=\thaap$, $\phii=\phiip$
of the $\Z_2$ orientifold action ---
we find
\be
\label{C2}
C_{\mu abc} = 2\phi_{\mu}\om_{abc} \,,
\ee
where $\om_{abc} = \sqrt{G_3}\ep_{abc} $ is the volume form
on $X_{3}$, with $\sqrt{G_3}$ given by eq. (\ref{X3det}).

In appendix B, we show that
$A^+_{\mu} = B_{\mu} - 16 \phi_{\mu}$
is the massless linear combination
proportional to the $\Ur$ gauge field
\be
\label{ARdef2}
A^+_{\mu} = -24\etatil  A^{R}_{\mu}\,,
\ee
where $A^{R}_{\mu}$ is the $\Ur$ gauge field
canonically coupled to the R-current.
Setting
$V^+_{\mu} = B_{\mu} +  8 \phi_{\mu} =0 $ we obtain
\be
\label{etatil}
\phi_\mu = \etatil A^{R}_{\mu}.
\ee
Using (\ref{C2}) and (\ref{etatil}) in (\ref{WZactrescale2}),
and integrating over $X_3$,
with $\mathrm{Vol}(X_3) = {16\pi^2}/{9}$,
we find
\be
\label{5dact2}
\etatil \frac{3 v}{16\pi^2}\int_{{\rm AdS}_5} \!\!
A^{R} \we [\tr (R\we R) +  4\, \tr (F\we F)]\,.
\ee
As in section \ref{N=2},
the non-invariance of this action under a
$\Ur$ gauge transformation implies an anomaly
\be
\label{stringanom2}
\langle \pa_{\mu} ( \sqrt{g} R^{\mu})\rangle_{\rm total}
= \,-\,\etatil \frac{3 v}{32\pi^2}  (R \tilde{R}  -  4\, F \tilde{F} )\,.
\ee
To match the field theory result (\ref{totalanom2}),
we must show that $\etatil = \frac{1}{12}$.

Before determining $\etatil$, however,
we will check that the relative normalization
between the two terms in eq.~(\ref{stringanom2})
works out correctly,
proceeding as in section \ref{N=2}.
The dimensional reduction of the gauge field kinetic term
in the D7-brane Born-Infeld action gives
\be \label{5dYMact2}
\la \mu_7 e^{-\Phi}(2\pi\al')^2 L^4 \mathrm{Vol}(X_3)
\int \D^{5}x \frac{1}{4}\sqrt{-g}\, F^a_{\mu\nu}F^{a\mu\nu}
\equiv
\frac{1}{4g_{\mathrm{F}}^2}
\int \D^{5}x \sqrt{-g}\, F^a_{\mu\nu}F^{a\mu\nu} \,.
\ee
This differs from the analysis in section \ref{N=2}
only in the change in $L^4$,
and in the volume of the intersection
of the D7-brane worldvolume with $T^{11}$, which we called $X_3$ above.
Using (\ref{Lconifold}) and $\mathrm{Vol}(X_3) = {16\pi^2}/{9}$,
we find ${1}/{g_{F}^2} = {3\la v}/{16\pi^2}$.

As in section \ref{N=2} we now compare
two different expressions for the correlator of two flavor currents.
{}From the AdS/CFT correspondence we find
\be
\label{JJJ1}
\langle J^a_{\mu}(x) J^{b}_{\nu}(0) \rangle =
{\de^{ab} \over 2\pi^2 g^2_{\mathrm{F}}}
(\eta_{\mu\nu} \Yfund -  \pa_{\mu}\pa_{\nu})
\frac{1}{x^4}
=
{3v \over  2(2\pi)^4}
\la  \de^{ab} (\eta_{\mu\nu} \Yfund -  \pa_{\mu}\pa_{\nu})
\frac{1}{x^4}\,.
\ee
The determination of the same correlator in the field theory is more subtle.
In general, this correlator has the form \cite{Anselmi:1998}
\be
\langle J^{a}_{\mu}(x)J^{b}_{\nu}(0)\rangle =
{{b}\over (2\pi)^4 }
\tr(t^at^b)
(  \eta_{\mu\nu} \Yfund - \pa_{\mu}\pa_{\nu} )\frac{1}{x^4}.
\ee
In the ultraviolet theory, $b_{\mathrm{UV}} = v$,
but we must find the value $b_{\mathrm{IR}}$
to which this flows in the infrared.
An exact non-perturbative equation
$b_{\mathrm{IR}} = (1-\ga_{\mathrm{IR}})b_{\mathrm{UV}}$
was derived in \cite{Anselmi:1998},
where $\ga_{\mathrm{IR}}$ is twice the
anomalous dimension of the bifundamentals.
Using $\ga_{\mathrm{IR}}=-\frac{1}{2}$ (see \eg{} \cite{Gubser:1998b})
yields  $b_{\mathrm{IR}} = \frac{3}{2}v$,
and therefore
\be
\label{JJJ2}
\langle J^a_{\mu}(x) J^{b}_{\nu}(0) \rangle =
{ 3v \over 2 (2\pi)^4 }
\tr(t^a t^b)
(\eta_{\mu\nu} \Yfund -  \pa_{\mu}\pa_{\nu}){1\over x^4}.
\ee
A comparison of (\ref{JJJ1}) and (\ref{JJJ2})
shows that the field theory and string theory normalizations
of $\tr(t^a t^b)$ coincide.

The final step is to determine the parameter $\etatil$.
The  supergravity action for the $\U(1)_R$ vector mode is obtained
from eq.~(\ref{5dredact2}) with $x=1$ together with eq.~(\ref{volT11}),
\be
\label{T11redact}
S = \frac{L^8}{2\ka^2}
\int_{T^{11}/ {\mathsf{Z}\kern -3.5pt \mathsf{Z}}_2 } \!\!\!\!\D^5y \sqrt{G}
\int_{\mathit{AdS}_5}\!\!\!\! \D^5 x \sqrt{-g}\, \frac{1}{3}
\left[- \frac{1}{4}  F(A^+)^2  \right]
\equiv -\frac{1}{4g_{\mathrm{SG}}^2}\int \D^5x \sqrt{-g} \,F(A^{R})^2\,.
\ee
Using (\ref{Lconifold}) and (\ref{ARdef2})
together with $\mathrm{Vol}(T^{11}/\Z_2) = {8\pi^3}/{27}$,
we find
${1}/{g_{\mathrm{SG}}^2} = {3^4 v^2\etatil^2}/{4\pi^2}$.

Next, we  compare two different calculations of the
correlator of two R-currents.
{}From the AdS/CFT correspondence, we have \cite{Freedman:1998}
\be
\label{RRR1}
\langle R_\mu(x)R_\nu(0) \rangle =
\,-\,\frac{1}{2\pi^2 g_{\mathrm{SG}}^2}
(\eta_{\mu\nu} \Yfund -  \pa_{\mu}\pa_{\nu})
\frac{1}{x^4}
=
\,-\, \frac{3^4 \etatil^2 v^2}{2^3\pi^4}
(\eta_{\mu\nu} \Yfund -  \pa_{\mu}\pa_{\nu})
\frac{1}{x^4}\,.
\ee
The field theory result for the same correlator is \cite{Anselmi:1997}
\be
\langle R_\mu(x)R_\nu(0) \rangle = \,-\, \frac{c}{3\pi^4}
(\eta_{\mu\nu} \Yfund -  \pa_{\mu}\pa_{\nu})
\frac{1}{x^4}\,,
\ee
where $c$ is the central charge,
related to the gravitational trace anomaly.
For the ultraviolet theory,
with $v(v+1)$ vector multiplets and $3v^2 + 9v$ chiral multiplets,
one may use the free field theory result \cite{Anselmi:1997}
$c_{\mathrm{UV}}= {\ts {1\over 8}} N_{\rm v} +{\ts {1\over 24}} N_{\chi}$
to find
$c_{\mathrm{UV}} ={1\over 4} v^2 $ to leading order in $N$.
The RG flow of the central charge satisfies \cite{Gubser:1998b,Anselmi:1998}
$c_{\mathrm{IR}}/ c_{\mathrm{UV}} =
\mathrm{Vol}(S^5/[\Z_2 {\times} \Z_2]) / \mathrm{Vol}(T^{11}/\Z_2)
= 27/32$,
thus in the IR fixed point theory, one has
\be
\label{RRR2}
\langle R_\mu(x)R_\nu(0) \rangle = \,-\, \frac{3^2 v^2}{2^7 \pi^4}
(\eta_{\mu\nu} \Yfund -  \pa_{\mu}\pa_{\nu})
\frac{1}{x^4}\,,
\ee
to leading order.
Comparing the results (\ref{RRR1}) and (\ref{RRR2}),
we find
$\etatil = \frac{1}{12}$.

\section{Summary} \label{summ}

In this paper we considered the $1/N$ correction to the $\U(1)$
R-current anomaly of a
broad class of models as tests of the AdS/CFT correspondence for string theory
on $\mathit{AdS}_5{\times} X_5$,
where either $X_5=S^5/G$ (where $G$ is the orientifold group),
or $X_5 = T^{11}/\Z_{2}$ (which arises from an orientifold of the conifold).
Whenever D7-branes and O7-planes are present,
the $1/N$ correction to the anomaly computed from the superconformal
field theory agrees with that obtained from string theory.
In models with no D7-branes or O7-planes,
the $1/N$ corrections to the anomaly are absent as well.
Our results give striking
confirmation of the Maldacena proposal at order $1/N$.

It would be interesting to pursue other examples of $1/N$ corrections
in the AdS/CFT correspondence, and to further understand
the $1/N^2$ corrections
(see \cite{Bilal:1999} for a discussion of $1/N^2$ corrections
to anomalies),
which in general involve string loops.

\section*{Acknowledgements}

We would like to thank Marc Grisaru for a useful discussion.

\section*{Appendices}

\appendix
\setcounter{equation}{0}
\section{The field theory gravitational $\UR$ anomaly} \label{Gravanom}

In this appendix, we calculate the gravitational $\UR$ anomaly
for the various classes of $\cN=2$ field theories considered in
sec.~2 of this paper.

The models arising from IIA theories with a pair of O$6^-$ planes
consist of $k$ NS5-branes on a circle, with $v_j$
D4-branes stretching between adjacent NS5-branes,
and $w_j$  D6-branes located between adjacent NS5-branes.
They fall into three classes \cite{Park:1998}:

\bigskip

\noindent (I.i)
$\Sp(v_0)\times \SU(v_1) \times \cdots \times \SU(v_P)$
with matter
$\bigoplus_{j=1}^{P} (\Yfund_{j-1},\Yfund_{j})
+ \Yasymm_{P} + \half   {w_0}  \Yfund_0 +
 \bigoplus_{j=1}^{P} w_j \Yfund_{j} $

\medskip
For an odd number $k=2P+1$ of NS5-branes,
one of the NS5-branes necessarily intersects an O$6^-$ plane.
The index $j$ ranges from $-P$ to $P$,
and the orientifold symmetry constrains
$v_{-j} = v_j$ and $w_{-j} = w_j$.
The requirement that the beta-function vanish for each gauge group
factor gives rise to the conditions (2.2) of ref.~\cite{Park:1998},
which are equivalent to
\bea
w_0 & = &2v_0 - 2v_1 + 4, \non\\
w_j & = &2v_j - v_{j-1} - v_{j+1}, \qquad 1 \le j \le  P-1   \\
w_P & = &v_P - v_{P-1} + 2. \non
\eea
The first $P$ equations are solved to give
\be
\label{Ii}
v_j = v_0 + 2j - \half j w_0 - \sum_{i=1}^{j-1} (j-i) w_i
\ee
and the last equation leads to the constraint
\be
\label{constraintIi}
\sum_{j= -P}^{P} w_j = 8.
\ee
The coefficient of the gravitational anomaly is proportional to
(see eq.~(\ref{Qdef}))
\bea
{Q \over 2} &=& \half v_0 (v_0 +1)  + \sum_{j=1}^P (v_j^2 -1)
- \sum_{j=1}^P v_{j-1} v_j - \half v_P (v_P-1) + \delta_{P,0}
- \half w_0 v_0 - \sum_{j=1}^P w_j v_j \non\\
&=& \half v_0 \left(  v_0 - v_1 - w_0 + 1 \right)
+ \half \sum_{j=1}^{P-1} v_j \left(2v_j - v_{j-1} - v_{j+1} -
2 w_j\right)\non\\
&&
+ \half v_P \left( v_P - v_{P-1} - 2 w_P  + 1\right)  - P
+ \delta_{P,0} \non\\
& =& - \half v_0 - \quarter v_0 w_0 - \half \sum_{j=1}^P v_j w_j - \half v_P
-P + \delta_{P,0}
\eea
where in the last line we used eq.~(\ref{Ii}).

Consider this theory in the large $N$ limit, where $N=\sum_j v_j$.
Equation (\ref{Ii}) implies that the $v_j$ are equal, to $\cO(N)$.
We denote this common value by $v= N/k$.
For a given model, $k$ is fixed so that $k \ll N$ in the large $N$ limit.
The apparent $\cO(N^2)$ contribution to the anomaly has cancelled out,
and the $\cO(N)$ contribution is
\be
{Q \over 2}
\sim v ( - 1- \quarter  w_0 - \half \sum_{j=1}^P  w_j)
= -3 v
\ee
where we have used eq. (\ref{constraintIi}).

\bigskip

\noindent ({I.ii})
$\Sp(v_0)\times \SU(v_1) \times \cdots \times \SU(v_{P-1}) \times \Sp(v_P)$
with matter
$\bigoplus_{j=1}^{P} (\Yfund_{j-1},\Yfund_{j})
+ \half w_0 \Yfund_0 + \bigoplus_{j=1}^{P-1} w_j \Yfund_{j}
+ \half w_P \Yfund_P
$

\medskip

In this class of theories, $k=2P$,
and none of the NS5-branes intersects an O$6^-$ plane.
The index $j$ ranges from $-P+1$ to $P$,
and the orientifold symmetry constrains
$v_{-j} = v_j$ and $w_{-j} = w_j$.
The requirement that the beta-function vanish for each gauge group
factor gives rise to the conditions (2.4) of ref.~\cite{Park:1998},
which are equivalent to
\bea
w_0 & = &2v_0 - 2v_1 + 4, \non\\
w_j & = &2v_j - v_{j-1} - v_{j+1}, \qquad 1 \le j \le  P-1   \\
w_P & = &2v_P - 2v_{P-1} + 4. \non
\eea
The first $P$ equations are solved to give
\be
\label{Iii}
v_j = v_0 + 2j - \half j w_0 - \sum_{i=1}^{j-1} (j-i) w_i
\ee
and the last equation leads to the constraint
\be
\label{constraintIii}
\sum_{j= -P+1}^{P} w_j = 8.
\ee
The coefficient of the gravitational anomaly is proportional to
\bea
{Q \over 2} &=& \half v_0 (v_0 +1)  + \sum_{j=1}^{P-1} (v_j^2 -1)
+\half v_P (v_P+1)
- \sum_{j=1}^P v_{j-1} v_j
- \half w_0 v_0 - \sum_{j=1}^{P-1} w_j v_j  - \half w_P v_P \non\\
& =& \half v_0 \left(  v_0 - v_1 - w_0 + 1 \right)
+\half \sum_{j=1}^{P-1} v_j \left(2v_j - v_{j-1} - v_{j+1}
- 2 w_j \right)\non\\
&& + \half v_P \left( v_P - v_{P-1} -  w_P  + 1\right) -P+1  \non\\
& = & - \half v_0 - \quarter v_0 w_0
- \half \sum_{j=1}^{P-1} v_j w_j
- \half v_P - \quarter v_P w_P -P+1
\eea
where in the last line we used eq.~(\ref{Iii}).
In the large $N$ limit,
the $\cO(N)$ contribution to the anomaly is
\be
{Q \over 2}
\sim v ( - 1- \quarter  w_0 - \half \sum_{j=1}^{P-1}  w_j -\quarter w_P)
= -3 v,
\ee
where we have used eq. ~(\ref{constraintIii}).

\bigskip
\noindent ({I.iii})
$\SU(v_1) \times \cdots \times \SU(v_P)$
with matter
$ \Yasymm_{1} +\bigoplus_{j=2}^{P} (\Yfund_{j-1},\Yfund_{j})
+\Yasymm_P + \bigoplus_{j=1}^{P} w_j \Yfund_{j}
$
\medskip

In this class of theories, $k=2P$,
and two  of the NS5-branes intersect an O$6^-$ plane.
The index $j$ ranges from $1$ to $2P$,
and the orientifold symmetry constrains
$v_{2P+1-j} = v_j$ and $w_{2P+1-j} = w_j$.
The requirement that the beta-function vanish for each gauge group
factor gives rise to the conditions (2.6) of ref.~\cite{Park:1998},
which are equivalent to
\bea
w_1 & = &v_1 - v_2 + 2, \non\\
w_j & = &2v_j - v_{j-1} - v_{j+1}, \qquad 2 \le j \le  P-1   \\
w_P & = &v_P - v_{P-1} + 2. \non
\eea
The first $P-1$ equations are solved to give
\be
\label{Iiii}
v_j = v_1 + 2j - 2 - \sum_{i=1}^{j-1} (j-i) w_i
\ee
and the last equation leads to the constraint
\be
\label{constraintIiii}
\sum_{j=1}^{2P} w_j = 8.
\ee
The coefficient of the gravitational anomaly is proportional to
\bea
{Q \over 2} &=&
\sum_{j=1}^P (v_j^2 -1)
- \half v_1 (v_1 -1) - \sum_{j=2}^{P} v_{j-1} v_{j} - \half v_P (v_P-1)
- \sum_{j=1}^P w_j v_j          \non\\
& =&\half v_1 \left(  v_1 - v_2 - 2 w_1 + 1 \right)
+ \half \sum_{j=2}^{P-1} v_j \left( 2v_j - v_{j-1} - v_{j+1} - 2 w_j \right)
\non\\
&& + \half v_P \left( v_P - v_{P-1} - 2 w_P  + 1\right) - P\non\\
& =& - \half v_1 - \half \sum_{j=1}^P w_j v_j - \half v_P-P
\eea
where in the last line we used eq.~(\ref{Iiii}).
In the large $N$ limit,
the $\cO(N)$ contribution to the anomaly is
\be
{Q \over 2}
\sim v ( - 1 - \half \sum_{j=1}^P  w_j)
= -3 v
\ee
where we have used eq. (\ref{constraintIiii}).

\bigskip
The models arising from IIA theories
with an O$6^+$ and an O$6^-$ plane
consist of $k$ NS5-branes on a circle, with $v_j$
D4-branes stretching between adjacent NS5-branes.
They fall into four  classes \cite{Park:1998}:

\bigskip
\noindent
({II.i})
$\Sp(v_0)\times \SU(v_1) \times \cdots \times \SU(v_P)$
with matter
$\bigoplus_{j=1}^{P} (\Yfund_{j-1},\Yfund_{j}) + \Ysymm_{P}$
\medskip

In this class of theories, $k=2P+1$,
and one of the NS5-branes intersects an O$6^-$ plane.
The index $j$ ranges from $-P$ to $P$,
and the orientifold symmetry constrains $v_{-j} = v_j$.
The requirement that the beta-function vanish for each gauge group
factor gives rise to the conditions (2.8) of ref.~\cite{Park:1998},
which implies
\be
\label{IIi}
v_j = v_0 + 2j.
\ee
The coefficient of the gravitational anomaly vanishes:
\bea
{Q \over 2} &=&
\half v_0 (v_0 +1)  + \sum_{j=1}^P (v_j^2 -1)
- \sum_{j=1}^P v_{j-1} v_j - \half v_P (v_P+1)
\non\\
& =& -P + \half v_0 \left(  v_0 - v_1 + 1 \right)
+ \half \sum_{j=1}^{P-1} v_j \left( 2v_j - v_{j-1} - v_{j+1} \right)
+ \half v_P \left( v_P - v_{P-1} -1 \right) \non\\
& =& -P - \half v_0 + \half v_P = 0
\eea
where we used eq. (\ref{IIi}) in the last line.

\bigskip
\noindent
({II.i$^\prime$})
$\SO(v_0)\times \SU(v_1) \times \cdots \times \SU(v_P) $
with matter
$\bigoplus_{j=1}^{P} (\Yfund_{j-1},\Yfund_{j}) + \Yasymm_{P}$
\medskip

In this class of theories, $k=2P+1$,
and one of the NS5-branes intersects an O$6^+$ plane.
The index $j$ ranges from $-P$ to $P$,
and the orientifold symmetry constrains $v_{-j} = v_j$.
The requirement that the beta-function vanish for each gauge group
factor gives rise to the conditions (2.10) of ref.~\cite{Park:1998},
which implies
\be
\label{IIiprime}
v_j = v_0 - 2j.
\ee
The coefficient of the gravitational anomaly vanishes:
\bea
{Q \over 2} &=& \half v_0 (v_0 -1)  + \sum_{j=1}^P (v_j^2 -1)
- \sum_{j=1}^P v_{j-1} v_j - \half v_P (v_P-1)
\non\\
& =& -P + \half v_0 \left(  v_0 - v_1 - 1 \right)
+ \half \sum_{j=1}^{P-1} v_j \left( 2v_j - v_{j-1} - v_{j+1} \right)
+ \half v_P \left( v_P - v_{P-1} + 1\right) \non\\
& =& -P + \half v_0 - \half v_P = 0
\eea
where we used eq. (\ref{IIiprime}) in the last line.

\bigskip
\noindent
({II.ii})
$\SO(v_0)\times \SU(v_1) \times \cdots \times \SU(v_{P-1}) \times \Sp(v_P) $
with matter
$\bigoplus_{j=1}^{P} (\Yfund_{j-1},\Yfund_{j})$
\medskip

In this class of theories, $k=2P$,
and none of the NS5-branes intersects either O$6$ plane.
The index $j$ ranges from $-P+1$ to $P$,
and the orientifold symmetry constrains $v_{-j} = v_j$.
The requirement that the beta-function vanish for each gauge group
factor gives rise to the conditions (2.12) of ref.~\cite{Park:1998},
which  implies
\be
\label{IIii}
v_j = v_0 - 2j.
\ee
The coefficient of the gravitational anomaly is proportional to
\bea
{Q \over 2} &=& \half v_0 (v_0 -1)  + \sum_{j=1}^{P-1} (v_j^2 -1)
+\half v_P (v_P+1)
- \sum_{j=1}^P v_{j-1} v_j
\non\\
& =& -(P-1) + \half v_0 \left(  v_0 - v_1 -  1 \right)
+ \half \sum_{j=1}^{P-1} v_j \left( 2v_j - v_{j-1} - v_{j+1}\right)
+ \half v_P \left( v_P - v_{P-1} + 1\right) \non\\
& = & -P+1 + \half v_0 - \half v_P = 1
\eea
where we used eq. (\ref{IIii}) in the last line.
This obviously has no $\cO(N)$ contribution.

\bigskip
\noindent
({II.iii})
$\SU(v_1)\times \cdots \times \SU(v_P) $
with matter
$ \Yasymm_{1} + \bigoplus_{j=2}^{P} (\Yfund_{j-1},\Yfund_{j}) + \Ysymm_{P} $
\medskip

In this class of theories, $k=2P$,
one of the NS5-branes intersects an O$6^-$ plane
and one of the NS5-branes intersects an O$6^+$ plane.
The index $j$ ranges from $1$ to $2P$,
and the orientifold symmetry constrains
$v_{2P+1-j} = v_j$ and $w_{2P+1-j} = w_j$.
The requirement that the beta-function vanish for each gauge group
factor gives rise to the conditions (2.14) of ref.~\cite{Park:1998},
which implies
\be
\label{IIiii}
v_j = v_1 + 2(j-1).
\ee
The coefficient of the gravitational anomaly is proportional to
\bea
{Q \over 2} &=&
\sum_{j=1}^P (v_j^2 -1)
- \half v_1 (v_1 -1) - \sum_{j=2}^{P} v_{j-1} v_{j} - \half v_P (v_P+1)
\non\\
& =& -P + \half v_1 \left(  v_1 - v_2 + 1 \right)
+ \half \sum_{j=2}^{P-1} v_j \left( 2v_j - v_{j-1} - v_{j+1} \right)
+ \half v_P \left( v_P - v_{P-1} - 1\right) \non\\
& =& -P - \half v_1 + \half v_P = -1
\eea
where we used eq. (\ref{IIiii}) in the last line.
This obviously has no $\cO(N)$ contribution.

\setcounter{equation}{0}
\section{KK reduction on $T^{11}$} \label{KK}

In this appendix, we discuss the
KK reduction of some of the massless fields
of IIB supergravity on $T^{11}$,
following closely the approach in ref.~\cite{Jatkar:1999}.
The authors of that paper observed that by Fourier expanding
all fields in the $\psi$ coordinate,
and transforming to the orthonormal tangent frame,
the covariant derivatives in this frame acting
on the Fourier-expanded fields are simply related to the
orthonormal frame derivatives of $S^2{\times}S^2$ in the presence of a
magnetic monopole field on each sphere.
The presence of the magnetic monopole field results
from the fact that $T^{11}$ is not a product manifold.

Before proceeding let us establish some conventions.
The f\"unfbein $e_a{}^{\ua}$
(where $a,b,\ldots$ label the coordinates of $T^{11}$
and $\ua,\ub,\ldots$ label an orthonormal frame in the tangent space)
is
\be
\label{funfbein}
e_a{}^{\ua}
= \pmatrix{ \third           & 0 & 0       & 0        & 0 \cr
          0                 & \rtsix  & 0       & 0        & 0 \cr
          \third  \cos\thaa & 0 & \rtsix\sin\thaa &0 & 0 \cr
          0                 & 0 & 0 & \rtsix & 0  \cr
          \third \cos\thaap & 0 & 0 & 0 &\rtsix \sin\thaap \cr} \,, \qquad
(a=\psi,\,\thaa,\,\phii,\,\thaap,\,\phiip)
\ee
and is related to the metric (\ref{T11met}) by
$g_{ab} = e_{a}{}^{\ua}\de_{\ua\ub}e_{b}{}^{\ub}$,
where $\de_{\ua\ub}$ is the flat euclidean metric.
The inverse f\"unfbein $e_{\ua}{}^{a}$
obeys $e_{\ua}{}^{a} e_{a}{}^{\ub} = \de_{\ua}{}^{\ub}$,
$e_{a}{}^{\ua} e_{\ua}{}^{b} = \de_{a}{}^{b}$,
and $g^{ab} = e_{\ua}{}^{a}\de^{\ua\ub}e_{\ub}{}^{b}$.

The tangent frame derivative is given by $D_{\ua} = e_{\ua}{}^aD_a$.
Acting on a scalar expanded in the $\psi$ coordinate,
$\sum_s \e^{is\psi} \Phi_s(\thaa, \phii, \thaap, \phiip)$,
one finds  \cite{Jatkar:1999}
\be \ba {rcll}
D_{\underline{5}}\Phi_s
&=&  3is  \Phi_s\,, &\non\\
D_{\underline{r}} \Phi_s
&=& (\pa_{\underline{r}} + e_{\underline{r}}{}^{5}\pa_{5} )\Phi_s
= (\pa_{\underline{r}} + is\om_{\underline{r}})\Phi_s
= \del_{\underline{r}}\Phi_s\,, & r = 1,2\,, \non\\
D_{\underline{\rpr}} \Phi_s
&=& (\pa_{\underline{\rpr}} + e_{\underline{\rpr}}{}^{5}\pa_{5} )\Phi_s
= (\pa_{\underline{\rpr}} + is\om_{\underline{\rpr}})\Phi_s
= \del_{\underline{\rpr}}\Phi_s\,, & \rpr = 1^\prime, 2^\prime \,.
\ea
\ee
Here we have labeled  the coordinate $\psi$ by 5, the
coordinates $(\thaa,\phii)$ by $1,2$ and
the coordinates $(\thaap,\phiip)$ by $1',2'$.
The orthonormal frame coordinates are similarly labeled
$\underline{5},\underline{1},\underline{2},\underline{1}',\underline{2}'$.

It is convenient to introduce the new basis
$\del_{\pm}
= \frac{1}{\sqrt{2}}(\del_{\underline{1}} \mp i \del_{\underline{2}})$ and
$\del_{\pm^\prime}
= \frac{1}{\sqrt{2}}(\del_{\underline{1^\prime}} \mp i
\del_{\underline{2^\prime}})$.
It is important that $\om_{\underline{r}} = -e_{\underline{r}}{}^5$ is
precisely the connection in the
$\underline{r}$ directions \cite{Jatkar:1999}.
{}From this result one finds, for example, that when
acting on the $\pm$ component of a vector,
one can use the same derivative ($\del$) provided that
one changes the monopole charge $s$ to $s\pm 1$.
We refer to~\cite{Jatkar:1999} for further details.

First consider the Laplacian acting on a scalar field,
\be
\Yfund \Phi = D^a D_a \Phi =
\left[
\del_{\underline{r}} \del_{\underline{r}}
 +\del_{\underline{\rpr}} \del_{\underline{\rpr}}
- 9 {s^2}
\right] \Phi.
\ee
The eigenfunctions are \cite{Gubser:1998b, Jatkar:1999}
\bea
\label{scalarlaplacian}
\Phi^{l ,\lpr}_{s,m,\mpr}
&=&
\e^{is\psi}  Y^{l}_{s,m}Y^{\lpr}_{s,\mpr} \,, \non\\
\Yfund \Phi^{l ,\lpr}_{s,m,\mpr}
&=&
-H_{0}(l,\lpr,s)\Phi^{l,\lpr}_{s,m,\mpr} \,, \non\\
H_0(l,\lpr,s)
&=&
6[l(l+1) -s^2] + 6[\lpr(\lpr+1) - s^2] + 9s^2 \,,
\eea
where the $Y^{l}_{s,m}$ are the monopole harmonics
(see \eg{} \cite{Wu:1976}).

Next we turn to the case of interest, namely the vector modes.
The massless modes arise from the
$g_{\mu a}$ component of the metric and the
$C_{\mu abc}$ component of the four-form.
We expand these fields as\footnote{There is also an additional
contribution \cite{Ceresole:1999a} to the expansion
of $C_{\mu abc}$ which gives rise to the so called
massless Betti vector, but that mode is not relevant
to our discussion since it couples to the baryon number current.}
$g_{\mu a}   = B^i_\mu     Y^i_{a} $ and
$C_{\mu abc} = \phi^i_\mu \sqrt{G} \ep_{abc}{}^{de}D_{d} Y^i_{e}$,
where $Y_{a} = e_a{}^{\ua} Y_{\ua}$,
with $Y^i_{\ua}$ the eigenvectors of the Laplace-Beltrami operator,
$D_{\ua}D_{\ua} - 4$,
in the $T^{11}$ directions.
The $Y^i_{\ua}$ can be expanded in monopole harmonics
on $S^2{\times} S^2$,
\be
Y^i_{\ua}
= \left( \ba{c} Y^i_{\underline{5}} \\
                  Y^i_+ \\ Y^i_- \\ Y^i_{+'} \\ Y^i_{-'} \ea\right)
\sim
\left( \ba{c}
\e^{is\psi}  Y^{l}_{s,m}Y^{\lpr}_{s,\mpr}\\
\e^{is\psi}  Y^{l}_{s+1,m}Y^{\lpr}_{s,\mpr}\\
\e^{is\psi}  Y^{l}_{s-1,m}Y^{\lpr}_{s,\mpr}\\
\e^{is\psi}  Y^{l}_{s,m}Y^{\lpr}_{s+1,\mpr}\\
\e^{is\psi}  Y^{l}_{s,m}Y^{\lpr}_{s-1,\mpr}
\ea\right)
\ee
The action of $\Yfund = D_{\ua}D_{\ua}$ on each of the components
has been worked out in \cite{Jatkar:1999};
using these results we find
\be
\label{vectormatrix}
(\Yfund -4)
\left( \ba{c} Y^i_{\underline{5}} \\ Y^i_+ \\
                    Y^i_- \\ Y^i_{+'} \\ Y^i_{-'} \ea\right)
= \left(\ba{ccccc} -H_0 - 8 & f(l,-s) &-f(l,s) &f(\lpr,-s) &-f(\lpr,s) \\
                f(l,-s)     & -H_0 +6s   & 0 &0&0 \\
               -f(l,s)      & 0 & -H_0 -6s&0 & 0 \\
                f(\lpr,-s) & 0 &0 &-H_0 +6s  & 0 \\
               -f(\lpr,s) &0 &0 &0 &-H_0 -6s
\ea \right)
\left( \ba{c} Y^i_{\underline{5}} \\ Y^i_+ \\
                    Y^i_- \\ Y^i_{+'} \\ Y^i_{-'} \ea\right)
\ee
where
\bea
H_0 &=&
H_0 (l, \lpr, s) ~=~ 6[l(l+1)  + \lpr(\lpr+1)] - 3 s^2 \,, \non\\
f(l,s) &=& \sqrt{12(l+s)(l-s+1)} \,.
\eea
The components of $Y^i_{\ua}$ are not independent;
this means that we will find the eigenvalue of the scalar
laplacian (\ref{scalarlaplacian})
among the eigenvalues of the above mass matrix \cite{Ceresole:1999a}.
Discarding this eigenvalue, the remaining eigenvalues of the mass matrix
are \cite{Ceresole:1999a}
\be
-3-H_0(l, \lpr, s\pm 1)\,;\quad
-H_0(l, \lpr, s)- 4 \mp 2 \sqrt{H_{0}(l, \lpr, s)+ 4}\,.
\ee

We will need not only the mass eigenvalues,
but also the action for the vector modes.
The reduction of the supergravity action on $S^5$ was
worked out in \cite{Arutyunov:1998}.
The reduction on $T^{11}$ is similar;
the only difference lies in the eigenvalues of the vector harmonics,
$(\Yfund -4) Y^i_{\ua} = \la Y^i_{\ua}$.
Starting from eq. (3.16) of ref.~\cite{Arutyunov:1998},
we can diagonalize the action by the change of basis
\bea
\label{newbasis}
A_{\mu} &=& B_{\mu} - 4(1 + \sqrt{1-\la})\phi_{\mu} \,, \non \\
V_{\mu} &=& B_{\mu} - 4(1 - \sqrt{1-\la})\phi_{\mu} \,.
\eea
We will concentrate on the eigenvalues
$\la_{\pm} = -H_0 - 4 \mp 2 \sqrt{H_0+4}$.
Writing $\sqrt{H_0 + 4} = 2 x$,
so that $\la_{\pm} = -4(x^2 \pm x)$ and
$\sqrt{1-\la_{\pm}} = 2(x\pm \frac{1}{2})$
(the choice of branch here involves no loss in generality),
the reduced supergravity action for the field $A^{\pm}_{\mu}$ becomes
\be
\label{5dredact2}
\frac{L^8}{2\ka^2}(\int\sqrt{G}G_{ab}Y^a Y^b)
\int \D^5 x\sqrt{-g}
 \frac{(x{-}\half{\pm}\half)} {2(x\pm \half)}
\Big[
         -{\ts \frac{1}{4}}F_{\mu\nu}(A^{\pm})^2
	- 2 (x{-}1)(x{-}1{\pm}1) A^{\pm}_{\mu}A^{\pm \mu}
\Big]\,.
\ee
Thus, the $A^{+}_{\mu}$ mode is massless when $x=1$,
corresponding to $H_0 = 0$, that is,  $l=\lpr=s=0$.
This massless vector couples to the $\Ur$ current.
(This axial vector
sits in the gravity supermultiplet,
together with two  gravitini and a graviton,
which is consistent with the fact that the $\Ur$ current sits in the same
multiplet as the stress tensor which couples to the graviton.
The $A^{-}_{\mu}$ mode is massless for $x=2$,
corresponding to $H_0 = 12$, that is,  $l=1, \lpr=s=0$
or $\lpr=1, l=s=0$.
These two massless $\SU(2)$ triplet vectors
sit in vector supermultiplets and are associated with
the two $\SU(2)$ factors of the isometry group.)
The $V^{\pm}_{\mu}$ modes  are massive.

Let us now concentrate on the mode corresponding to the $\Ur$ current.
For $x=1$, we have $\la_+ = -8$,
so that equation (\ref{newbasis}) becomes
\bea
\label{newbasisplus}
A^+_{\mu} &=& B_{\mu} - 16 \phi_{\mu} \,, \non \\
V^+_{\mu} &=& B_{\mu} +  8 \phi_{\mu} \,,
\eea
the same result as for the $S^5$ case
(which follows from the fact that the eigenvalue
of the Laplace-Beltrami operator
is the same).

Finally, we discuss the form of the vector harmonic for this mode.
Setting $l=\lpr=s=0$,
the matrix (\ref{vectormatrix})
has only one non-zero entry,
namely $-8$ in the upper left hand corner.
The eigenvector corresponding to $\la = -8$ is
$Y_{\ua} = (Y_{\underline{5}},0,0,0,0)^T$,
where $Y_{\underline{5}}$
is independent of the angular variables for $l=\lpr =s=0$.
We will take $Y_{\ua} = (1,0,0,0,0)^T$.
Hence, for this mode
\be
\label{volT11}
\int \D^5 y \sqrt{G}G_{ab}Y^{a}Y^{b}
= \int \D^5 y \sqrt{G}\de_{\ua\ub}Y_{\ua}Y_{\ub}
= \int \D^5y \sqrt{G}.
\ee
Using the f\"unfbein (\ref{funfbein}),
one then obtains the vector harmonic for the $\Ur$ mode
\be
\label{vectorhmc}
Y_{a} = e_a{}^{\ua} Y_{\ua}
= (\third, 0 , \third  \cos\thaa , 0, \third \cos\thaap )^T.
\ee

\begingroup\raggedright\endgroup

\end{document}